# GHz sample excitation at the ALBA-PEEM

Muhammad Waqas Khaliq[1,2*], José M. Álvarez[1], Antonio Camps[1], Nahikari González[1], José Ferrer[1], Ana Martinez-Carboneres[1], Jordi Prat[1], Sandra Ruiz-Gómez[3], Miguel Angel Niño[1], Ferran Macià[2,4], Lucia Aballe[1], and Michael Foerster[1*]

[1] ALBA Synchrotron Light Facility, Carrer de la Llum, 2 – 26, 08290 Cerdanyola del Valles, Barcelona, Spain.

[2] Department of Condensed Matter Physics, University of Barcelona, 08028 Barcelona, Spain.

[3] Max Planck Institute for Chemical Physics of Solids, Noethnitzer Str. 40, 01187 Dresden, Germany.

[4] Institute of Nanoscience and Nanotechnology (IN2UB), University of Barcelona, 08028 Barcelona, Spain.

**Abstract**

We describe a setup that is used for high-frequency electrical sample excitation in a cathode lens electron microscope with the sample stage at high voltage as used in many synchrotron light sources. Electrical signals are transmitted by dedicated high-frequency components to the printed circuit board supporting the sample. Sub-miniature push-on connectors (SMP) are used to realize the connection in the ultra-high vacuum chamber, bypassing the standard feedthrough. A bandwidth up to 4 GHz with -6 dB attenuation was measured at the sample position, which allows to apply sub-nanosecond pulses. We describe different electronic sample excitation schemes and demonstrate a spatial resolution of 56 nm employing the new setup.

*Corresponding authors.

*E-mail address:* mkhaliq@cells.es (M. Waqas Khaliq), mfoerster@cells.es (M. Foerster)

# Introduction

Photoemission electron microscopy (PEEM) is a powerful surface characterization technique, especially when coupled to a synchrotron X-ray beamline (XPEEM mode) in order to excite electrons from the sample [1]. Many instruments in use around the world are based on a low energy electron microscope (LEEM) setup [2], where the sample stage is held at negative high voltage (extractor voltage) with respect to the cathode lens (at ground potential). The accelerating voltage is a requisite for the electron microscope operation, where the image resolution depends on the accelerating electric field. Although this scheme is the most common, other instruments keep the sample at ground potential and the electron optics at high positive potential to perform the imaging [3, 4]. For this other type of instruments, i.e., sample stage at ground potential, electrical connections to the stage are arguably easier and high-frequency (HF) options have been already reported [5].

The PEEM at the CIRCE beamline of the ALBA Synchrotron is based on an Elmitec LEEM-III microscope with sample stage at high voltage and is in operation since 2012 [6]. The standard four electrical contacts to the sample in this instrument are normally used for heating and temperature readout (thermocouple) and are not optimized for high frequency signal up to the GHz range. The ALBA setup disposes of two additional DC contacts, often used for electric sample excitation [7]. All electrical feedthroughs into the ultra-high vacuum (UHV), are integrated into a ceramic block permanently welded onto the sample stage. Although the external cables running from the supply rack to the instrument were upgraded to coaxial types [7], the UHV feedthroughs and the signal path inside the sample holder cartridge itself are not impedance-matched with the coaxial cables, and act as a bottleneck for the high frequency signal transmission.

There are two common experiments requiring high-frequency contacts to the sample: short (few nanoseconds and below) electrical pulses, and high-frequency continuous excitation. Short pulses are used either for static (before-after) or for time-resolved (pump-probe) measurements [8 – 11] generated either directly by pulse generators or from optically triggered switches close to the sample, while continuous signals are typically restricted to time- or phase- resolved experiments. For example, in order to observe current induced domain wall motion in magnetic nanowires [12, 13] or Skyrmion motion in flat stripes [14], short high-current density pulses were applied in XPEEM using a setup similar to Ref. [15]. When comparing images taken before and after the pulse, shorter pulses permit a more precise measurement of the motion velocity because imperfections in the samples can act as pinning sites and lead to a systematic underestimation of the speed [12]. Even shorter current pulses are available in other techniques, for example, the generation and the dynamics of skyrmions has been illustrated with the nanosecond current pulses in symmetric bilayers employing AFM [16], and in chiral-lattice magnet by Lorentz Transmission Electron Microscope [17]. Time resolved STXM has been used to visualize the Skyrmion Hall effect [18]. Additionally, the orbit of a magnetic vortex under current excitation has been observed by ultrafast Lorentz microscopy [19].

Examples of continuous excitation, using periodical HF signals, are XPEEM experiments with surface acoustic waves (SAW) [20-24]. Some experiments are still possible using the standard feedthroughs up to 500 MHz, matching the ALBA X-ray repetition rate. Lower sub harmonics (125 MHz, 250 MHz) can be accessed in combination with a gating system in the electron beam path [20]. Although at 500 MHz, only few percent of the original RF power arrives to the sample, effects driven by the dynamic strain of the SAW like magnetic domain wall motion and their delay [21, 22], magneto-acoustic waves [23] and work function oscillations in Pt [24] have been observed. However, dynamical effects at higher frequencies, like the intrinsic ferromagnetic resonance (FMR), have remained elusive due to the continued fast decay of transmission in that range [7].

In this manuscript we describe the design and operation of a high-frequency setup using additional cabling and bypassing the standard feedthrough in order to allow the use of dedicated high-frequency compatible components all the way to the printed circuit boards (PCBs) supporting the sample. The bandwidth extends well into the GHz range. In order to mitigate the pertinent issue of high-voltage arcs between sample and cathode lens, posing any electronics connected to the sample stage at risk of damage, modular and comparatively low-cost electronics are used. A spatial resolution of 56 nm has been demonstrated using the new setup and a reduced electric accelerating field (large distance between the objective lens and sample), as is typically used for this type of experiments. Furthermore, this work has also been granted as a patent in Spain [25].

## I. High-Frequency (HF) Setup

The new setup has been developed to be compatible with the existing XPEEM station at the CIRCE Beamline at the ALBA Synchrotron Facility. It constitutes a modular addition, i.e., it can be separately mounted and dismounted without breaking the chamber vacuum using a pumpable gate valve. A single permanent modification of the existing setup was the removal of a mu-metal shield around the objective lens. The setup comprises additional coaxial cables and SMP plugs to realize a connection to a printed circuit board integrated into the sample holder (Fig. 1), all made with UHV compatible materials. The design details and electrical characteristics are described below.

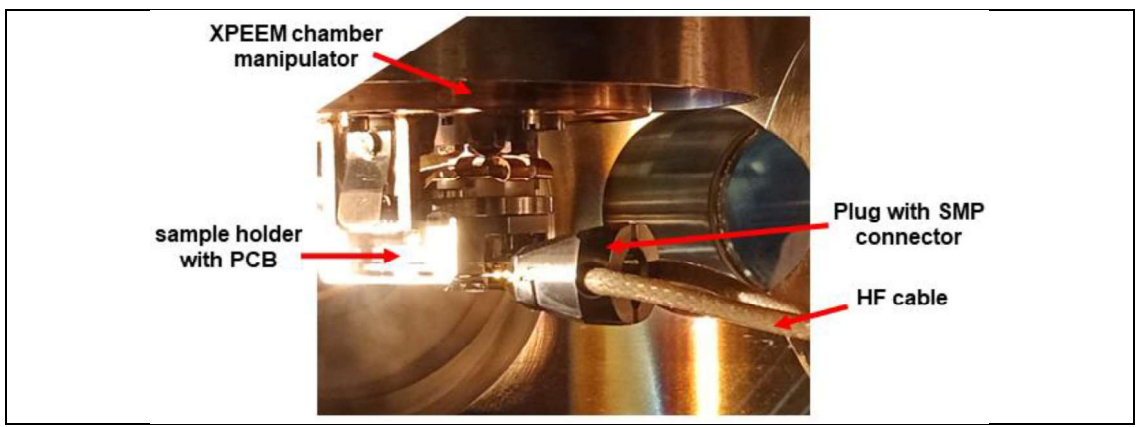

**Figure 1.** Connection scheme between the sample holder and the high frequency cable inside the XPEEM chamber. The cable provides a signal path to the sample through a plug that contains two SMP connectors.

### A. HF Cable and Plug

A schematic illustration of the HF setup together with pictures of the designed set up is depicted in Fig. 2. It is a long Y-shaped assembly using CF 40 UHV components. As can be seen in the picture, the opening on one side contains the electrical vacuum feedthrough (Sub-Miniature version A, SMA, standard) while the other is fitted with a wobble stick that serves to engage and disengage the plug, which connects the cables with the sample holder. The connection between feedthrough and plug is made by coaxial Teflon isolated cables (Allectra, model: 312-PTFE50-S). A thin (1.8 mm) version was selected due to its better flexibility despite having slightly higher electrical losses than the thicker versions. The central section of the setup comprises a vacuum bellow, which compensates the travel of the setup between the retracted position (behind the gate valve) and the position in contact with the sample stage. In this way the in-vacuum cabling does not fold or stretch, a critical point since the cable reference potential is the high voltage of the sample stage, while the outer assembly is at ground potential. A set of polyether ether keton (PEEK) insulators is placed inside the new setup at various points to further avoid the contact of the HF cable with the inner metallic walls of the system. On the other hand, the SMA feedthrough itself is also at high-voltage potential and electrically separated from the rest by a ceramic break.

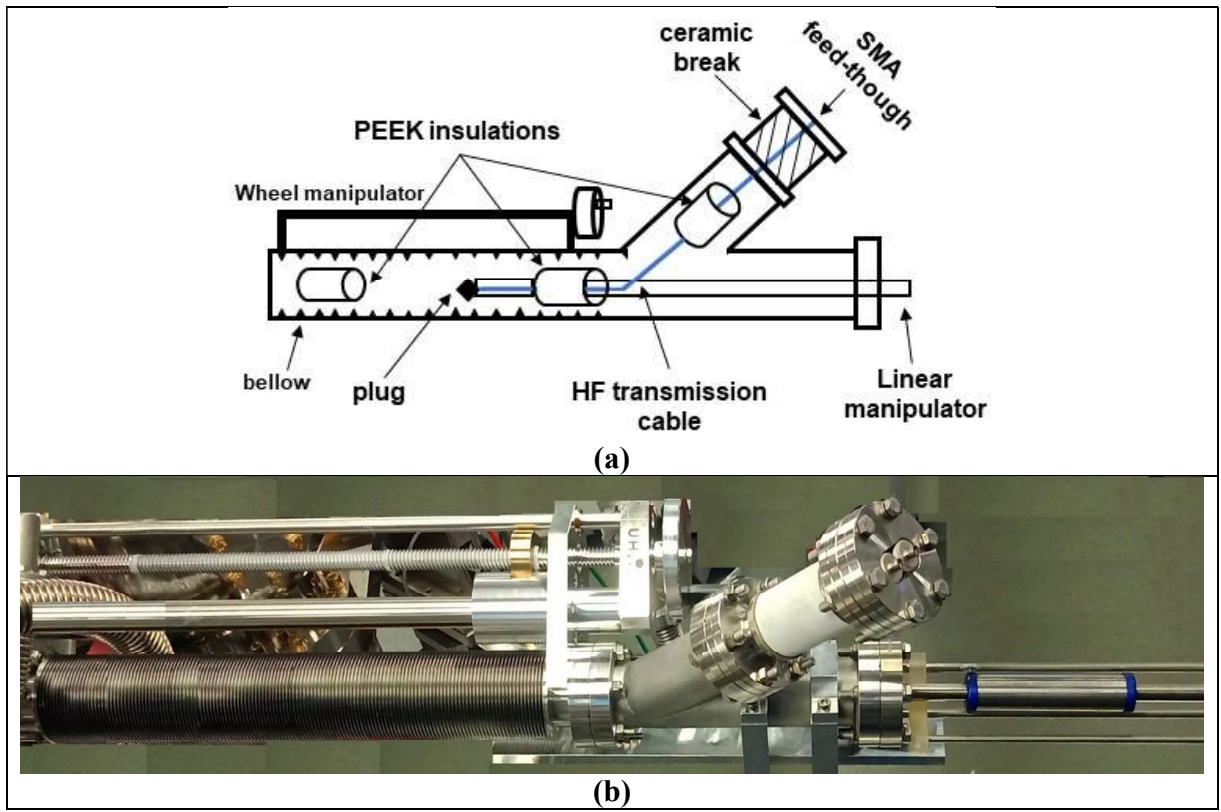

(a)

(b)

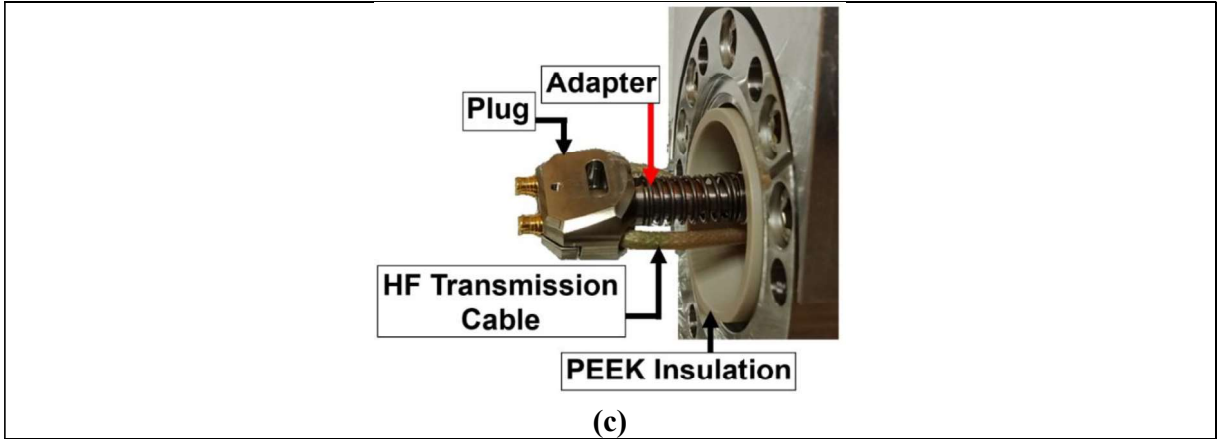

(c)

**Figure 2a.** Schematic illustration of the HF setup depicting the various components **b.** real view, **c.** side-view of the plug connected to linear manipulator through an adapter. The plug incorporates SMP connectors which link the cable to the printed circuit board in the sample holder.

### B. Sample Holder and Compatible PCB Design

In order to transmit the HF signal to the sample, dedicated PCBs compatible with UHV have been designed (See Fig. 3a). The SMP male connectors are directly soldered to the PCB. Samples are glued by silver paste onto the central area of the PCB and electrical contacts realized with ultrasonic wire bonds. The SMP counterpart is integrated into the plug at the end of the cable, shown in Fig. 2c. Figure 3b shows a combined image of the sample holder with a mounted sample, without (left) and with (right) sample holder cap. The modified cap has cut-outs for the bulky SMP connectors and a central hole for imaging. Its lower surface has to be separated from the sample surface to avoid short circuits with the wire bonds, the same procedure as for low-frequency electrical sample connections [7]. The sample holder is compatible with the incorporation of a small electromagnet below the PCB.

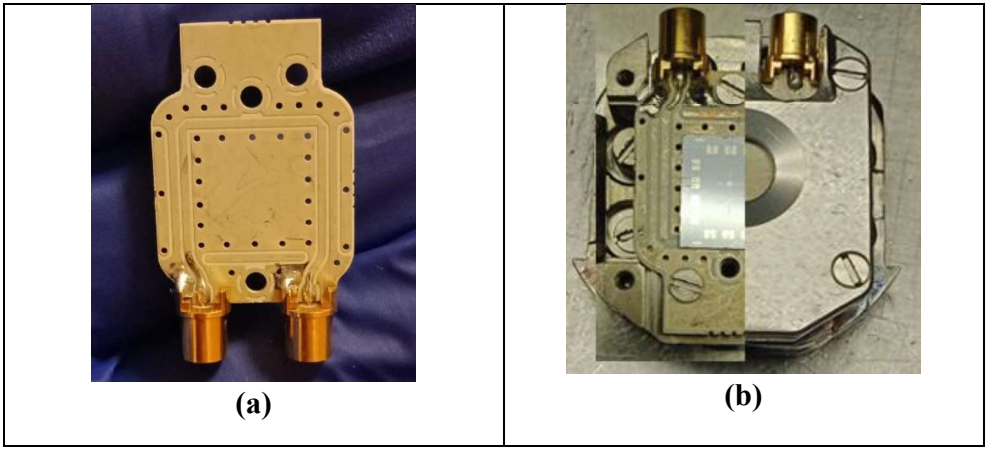

(a)          (b)

**Figure 3a.** PCB soldered with SMP connector through which the signal is transmitted to the sample, and **b.** Sample mounted on the PCB in the sample holder.

### C. Electrical Characterization

Figure 4 shows the electrical characterization of the HF connection in blue color compared to the old setup in black (standard vacuum feedthrough and sample holder, but with the same HF PCB and with coaxial cables on the air side). In panel a) of Fig. 4, the transmitted signals ($S_{12}$) measured with a network analyzer (Keysight FieldFox N9913A with -6 dBm excitation power) are shown. For these measurements, the transmitted signal is passing directly through the coplanar electrode on the PCB (no sample or wirebonds) and includes two coaxial connections which can be used, for example, as an in and output of a signal (transmission). Thus, the signal attenuation at the sample level corresponds to about one half of the total attenuation of $S_{12}$. $S_{ij}$ is defined as the ratio of signal power measured in port $i$ of a network analyzer vs the signal power sourced from port $j$. This means that $S_{12}$ corresponds to the transmission through our setup, where port 1 and port 2 are connected to two coaxial cable system. Both measurements were taken including the high voltage cable (depicted in figure 5 with black meshed structure) between the PEEM main HV rack and the microscope itself. While the old setup shows strong attenuation already in the MHz range, the useful bandwidth of the new setup is strongly improved, reaching into the low GHz range. We would like to stress that this is achieved by combining standard of-the-shelf HF components into an assembly compatible with the PEEM instrument operating conditions (HV and UHV). As an application example, in the inset of Fig. 4a the measurement of a $LiNbO_3$ sample with a SAW antenna and receiver (opposing interdigital transducers, IDTs) is shown. The characteristics of the sample with transmission peaks at the resonance frequencies of the IDTs are clearly visible. No comparable data can be obtained with the old setup, despite the fact that we have used it successfully in the past for experiments with SAW from 125 to 500 MHz. Concerning short electrical pulses, Fig. 4b depicts measurements performed with a pulse generator (as discussed in the next section) and oscilloscope. For a pulse of about 1 ns width, the new setup provides clearly improved transmission, although it still causes some attenuation and peak broadening. The results of the electrical characterization are summarized in Table 1.

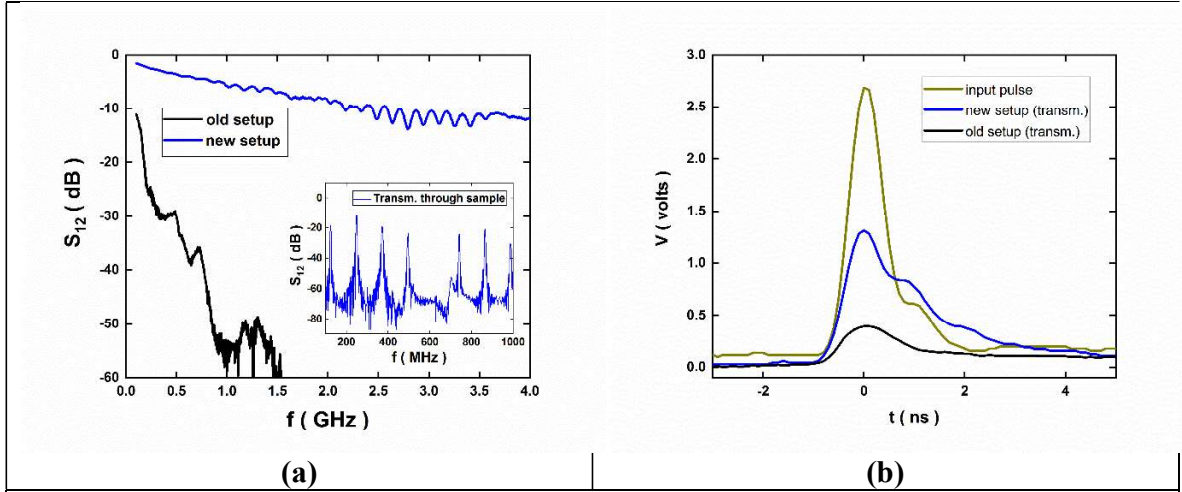

(a)                          (b)

**Figure 4a.** HF signal transmission through the new (blue) and old (black) setup. Inset: characterization of a $LiNbO_3$ sample SAW transmission filter inside the PEEM sample stage. **b.** Output comparison of transmitted nanosecond current pulses through a 50 Ohm load for both setups.

| Signal | New Setup $S_{12}$ (dB) | Standard feedthrough with RG 178 cable in air $S_{12}$ (dB) |
|---|---|---|
| 0.5 GHz | -3.6 | -30.0 |
| 1.0 GHz | -5.6 | N/A |
| 4.0 GHz | -11.9 | N/A |
| 1 ns pulse | -6.2 | -16.3 |

**Table 1.** Comparison of the signal attenuation in transmission through old and new setups. All measurements include cables from the rack to the microscope and use the same PCB as load.

## II. Electronic Equipment

Two different modular electronic systems are utilized to provide the high-frequency electric signals in the PEEM HV environment. Due to the pertinent risk of arcs in the PEEM microscope, which may cause damage to the electronics connected to the sample, a modular approach comprising simple electronic equipment has been adopted. Any connections from outside environment like a master clock into the HV rack need to provide galvanic separation of at least 10 kV and are therefore typically realized by optical fibers. The systems for synchronized sinusoidal signals and for single nanosecond pulses are illustrated in Figs. 5a and 5b, respectively.

The first system (see Fig. 5a) uses a Keysight EXG Vector signal generator model N5172B (RF generator) to capture the synchrotron master clock from a digital signal distributed to all ALBA beamlines by optical fiber [26, 27]. The RF generator produces a phase-synchronized analog sinusoidal signals with adjustable phase and power between 1.115 MHz (ALBA orbit clock) and 1 GHz. The ALBA photon bunch repetition rate is around 500 MHz, determined by the radio frequency accelerating cavities. The analog signal from the RF generator is transmitted via an optical fiber link designed at ALBA [28] into the PEEM high-voltage rack, where it is converted back to an electrical signal. Depending on the desired signal, different treatment such as pre-amplification, filtering and frequency multiplication is performed at low level before the final, universal high-power amplifier (Mini-Circuits ZHL-5W-422+). For example, in order to generate a 3 GHz signal, a starting frequency of 750 MHz is used, which is then amplified and doubled twice, with bandpass filters included after each doubling.

The system for applying short current pulses is shown in Fig. 5b. In its core, an AVTECH AVI-MP-P pulse generator provides short pulses (2-100 ns, 40 V) at an exceptional cost factor, at the expense of some flexibility (unipolar, fixed amplitude, and pulse width determined by the length of attached cable). The pulse generator is followed by either a fixed or a remote programmable attenuator (Mini-Circuits RCDAT-3000-63W2) which is controlled from a computer through an optical fiber. The pulse generator is controlled by an external trigger (here Keysight function generator 33220A) and needs an external power supply. The pulse transmitted through the sample can be analyzed by an oscilloscope.

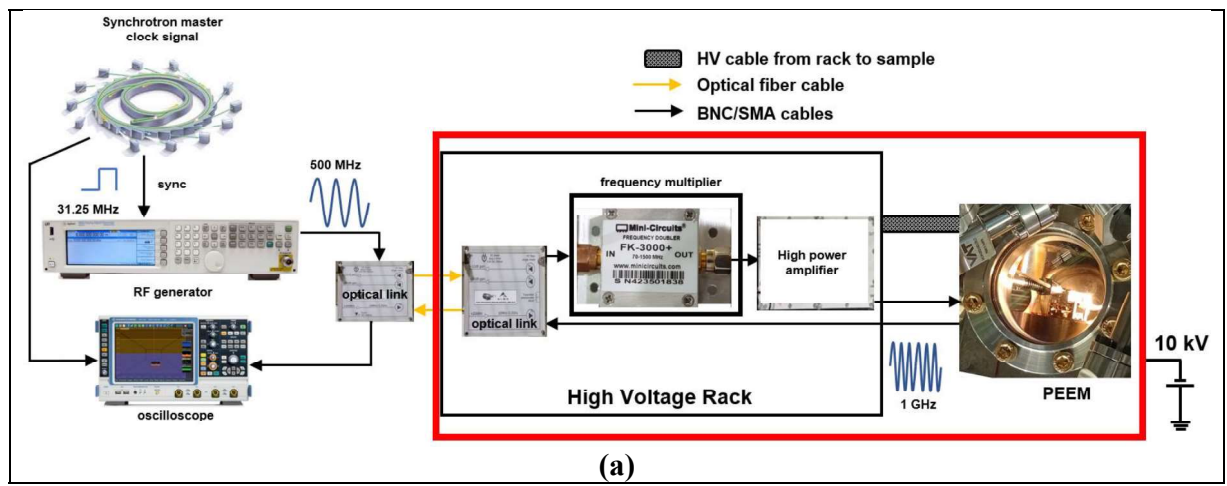

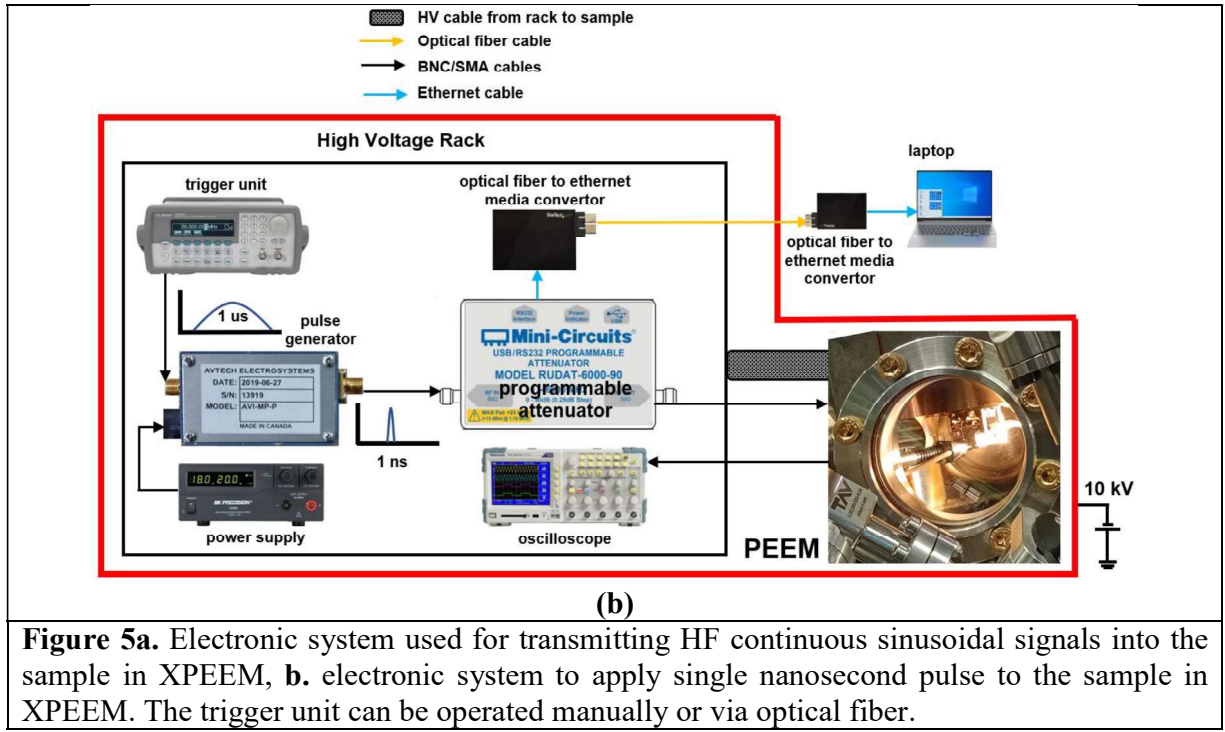

**Figure 5a.** Electronic system used for transmitting HF continuous sinusoidal signals into the sample in XPEEM, **b.** electronic system to apply single nanosecond pulse to the sample in XPEEM. The trigger unit can be operated manually or via optical fiber.

### III. Measurement examples and spatial resolution

As an example of data obtained using the high-frequency connection we present here the excitation of SAW of different frequencies in $LiNbO_3$. SAW show a strong and detectable XPEEM signal in phase-synchronized measurements [29]. These waves are generated by interdigitated transducer electrodes (IDT) at the edge of the sample and can travel several millimeters until the PEEM field of view where the measurements are performed. We focus here on the SAW direct electric contrast in the $LiNbO_3$ substrate, i.e. the oscillation of the local surface potential. Figures 6a and 6b show images of SAW at 1 GHz and 3 GHz with associated wavelengths of 4 um and 1.3 um, respectively. The electronic setup depicted in Fig. 5a was used, without/with frequency multiplier to yield a 1/3 GHz. 3 GHz signal is obtained by 2 steps of frequency doubling from 750 MHz signal. In order to measure the SAW electric amplitude, image sequences with varying sample bias voltage were acquired, which corresponds to the nominal kinetic energy of the detected photoemitted electrons. In particular, we scan around the low-energy photoelectron cutoff. For insulating samples, such as $LiNbO_3$, a local shift of the sample surface potential, e.g. due to the SAW, will result in a shift of the full electron spectrum, visible as spatial intensity modulations in XPEEM images [24]. The spectra extracted for red and blue boxes depicted in Fig. 6a and 6b are plotted in Fig. 6c and 6d, showing a 1 V amplitude for 1 GHz and a smaller 0.09 V for 3 GHz, mainly due to the decreasing IDT efficiency (due to the nonlinear dispersion relation of $LiNbO_3$, the second harmonic at 3 GHz is no longer tuned to an exact multiple of the 500 MHz and thus the IDT has to be operated at the limit of its resonance).

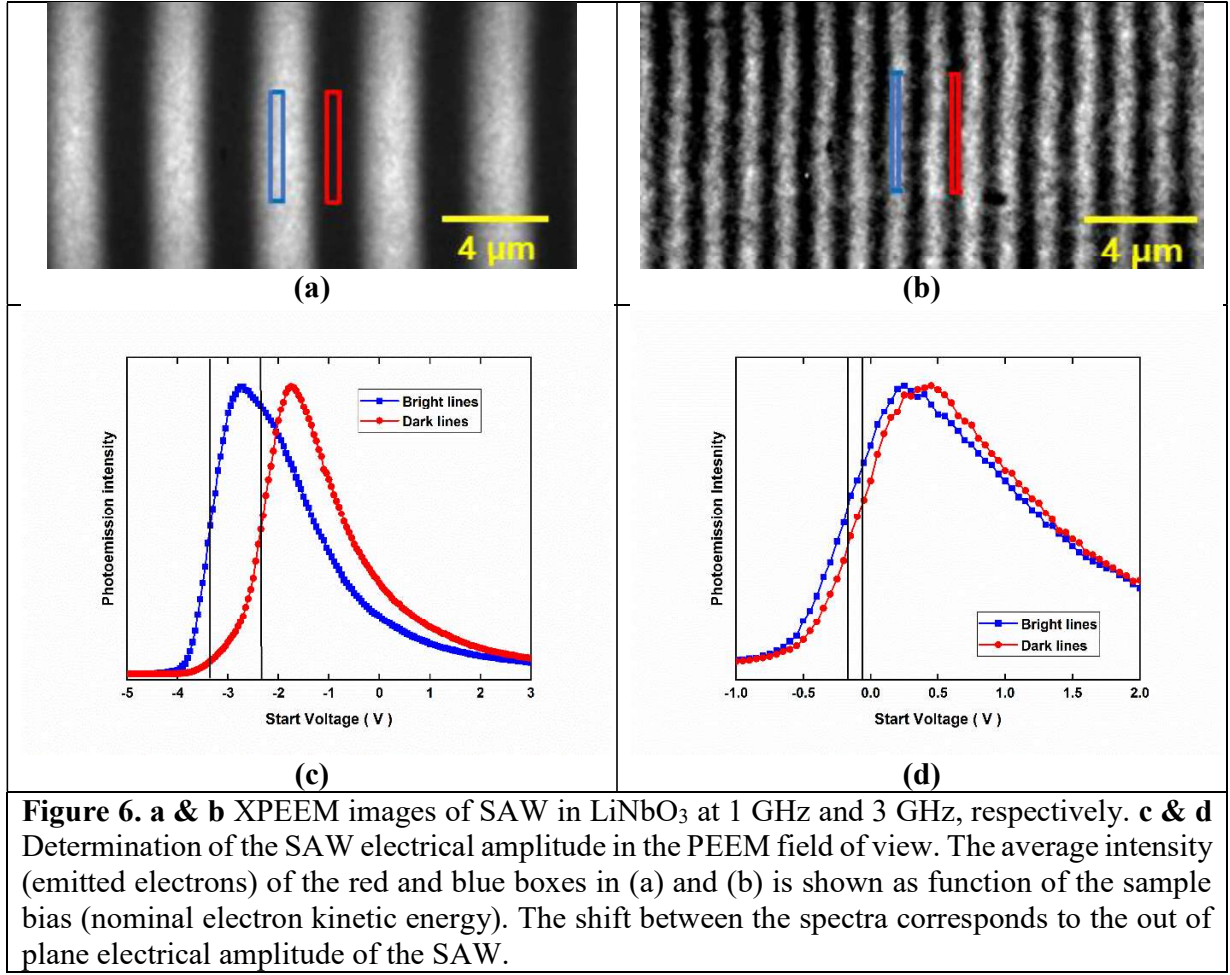

**Figure 6. a & b** XPEEM images of SAW in LiNbO$_3$ at 1 GHz and 3 GHz, respectively. **c & d** Determination of the SAW electrical amplitude in the PEEM field of view. The average intensity (emitted electrons) of the red and blue boxes in (a) and (b) is shown as function of the sample bias (nominal electron kinetic energy). The shift between the spectra corresponds to the out of plane electrical amplitude of the SAW.

We now analyze the performance of the new setup in terms of spatial resolution. The ultimate spatial resolution in XPEEM has been reported in the 20-30 nanometer range [6] and can be limited by several factors: quality of the electron optics, apertures and alignment, mechanical stability of the sample, detector sensitivity as well as the space charge of the electron bunches travelling inside the microscope [30, 31]. In practice, experiments involving electrical contacts on the sample surface such as electrodes or IDTs, typically use a reduced accelerating field to avoid arcs. The reasoning for this compromise in image quality is twofold; on the one hand the inclusion of electrical contacts on the sample surface and the gap between the cap and the sample could increase the risk of arcs for a given setting. On the other hand, electrode structures on the sample are typically destroyed by the very first arc, making it even more crucial than for homogenous samples to avoid arcs. The accelerating field is usually reduced by a factor 3-4 with respect to high-resolution imaging in order to reduce the risks of arcs between sample holder and objective lens. This is done by reducing the high voltage to 10 kV and increasing the sample distance from objective lens. On the one hand, arcs are more likely for many functional sample environments due to non-flat surfaces, insulating regions and/or degassing elements such as small electromagnets. On the other hand, contacted electrode structures on the sample are often destroyed by the very first arc event, probably due to the transient potential difference between different areas of the sample, which gives rise to equilibrating currents. Figure 7a shows an XPEEM image of a defect in a Co layer on a Si substrate, measured with the high frequency cable connected, at 10 kV operating voltage, 60% increased sample-objective distance and a floating cap; which are the typical parameters for experiments with active electrodes on the sample surface. The spatial resolution defined as 15-85% edge jump was determined to be 56 nm from the line profile in Fig. 7b and was equal to the one obtained without the HF cable on the same day.

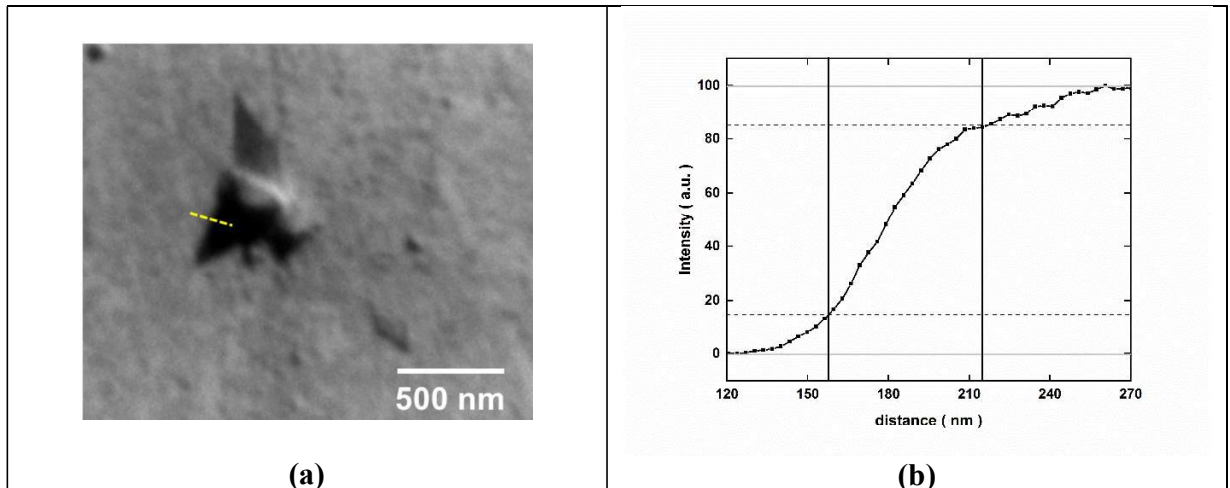

(a) (b)

**Figure 7a**. XPEEM image of a defect on a Co/Si sample at reduced acceleration voltage (10 kV) and increased sample objective distance; typical for experiments with injected electrical signals, taken at 777.8 eV photon energy and 1 V start voltage. **b.** Profile along the dashed yellow line in (a). The width of the grey box is 56 nm.

**Conclusions**

We presented a new setup to provide high-frequency electrical signals to the high-voltage sample stage of the ALBA XPEEM. The setup can be mounted and dismounted without modification of the main system or breaking the vacuum. The two lateral cable connections terminate in SMP connectors making a contact with counterparts at the PCB integrated in the sample holder. The measured bandwidth extends up to 4 GHz (-5.9 dB for one pass). Modular electronic systems are used for the generation of continuous and pulsed signals. A lateral resolution of 56 nm has been demonstrated in a routine measurement with typical settings for experiments employing on-sample electrodes. The setup, together with the recently installed electron gating system [10], provides excellent conditions for time resolved XPEEM measurements, for example to study the interaction of magnetic systems with high frequency SAW [32–35]. In addition, the application of shorter current pulses is now possible in a static before-after scheme which allows, for example, to study the effect of these pulses to magnetic domain wall or skyrmion motion.

**Acknowledgements**

We thank Bernat Molas and Bern Saló i Nevado from the ALBA electronics for their support in designing the PCBs and UHV fine soldering of SMP connectors to PCBs, respectively. Also, Alberto Hernández-Mínguez from Paul-Drude-Institut für Festkörperelektronik for providing lithium niobate samples with IDTs, prepared by Paul Seidel. We gratefully acknowledge the advice and help from the ALBA safety group, namely Ines Perez Renart, Jose Maria Roca, and Carme Mármol Moreno. The authors are also thankful to Spanish Ministry of Science, Innovation, Universities, and DOC-FAM who have received funding from the European Union's Horizon 2020 research and innovation program under the Marie Skłodowska-Curie grant agreement No. 754397. M.W.K., M.F., and M.A. Niño acknowledge the funding from MICINN through grant numbers RTI2018-095303-B-C53 and PID2021-122980OB-C54. F.M. is grateful to funding from MCIN/AEI/10.13039/501100011033 through grant number: PID2020-113024GB-100. This work has also been supported by the ALBA in-House Research Program.